# Physical mechanism of reconnection onset in space plasmas

M. Ugai

Research Center for Space and Cosmic Evolution, Ehime University,
Matsuyama 790-8577, Japan

(ugai@cosmos.ehime-u.ac.jp)

Magnetic reconnection plays a crucial role in large dissipative events in space plasmas, such as solar flares and geomagnetic substorms, but there has been much confusion regarding reconnection onset responsible for the explosive events. The main theme of the present paper is to propose a new physical concept on reconnection onset and to clarify physical mechanism responsible for flares and substorms. Reconnection rate can be defined as inductive electric field along X neutral line, so it is proposed that large-scale reconnection can occur when ambient reconnection (inductive) electric fields effectively penetrate into X line. In the solar corona or geomagnetic tail without effective Coulomb collision, reconnection electric fields are shielded (cut off) around X line, so reconnection cannot occur and magnetic energy can be stored in the form of current sheets. It is demonstrated that when plasmas become effectively collisional and dissipative around X line, shielding of reconnection electric fields around X line is violated, and reconnection drastically grows to eventually cause the (Petschek-like) fast reconnection mechanism responsible for flares and substorms.

## I. INTRODUCTION

In space plasmas, there are observed large dissipative events, such as solar flares in the solar corona and geomagnetic substorms in the geomagnetic tail.[1] These explosive events bring about drastic changes in large-scale plasma circumstances, so it is essential for space plasma studies to clarify physical mechanism of flares and substorms. In both flares and substorms, critically stored magnetic energy is explosively released into plasma energies, so the following two questions are fundamental. The first question is how magnetic energy can be stored in large-scale region. In this growth phase, effective dissipation of magnetic energy should not occur. The second question is how stored magnetic energy can be explosively released. In this explosive phase, physical mechanism of explosive magnetic energy release should be realized and work. Any reasonable theoretical model should explain these key questions self-consistently.

It is well recognized, both theoretically and observationally, that so-called magnetic reconnection plays an essential role in flares and substorms.[2] Historically, so-called tearing instabilities were found to occur locally in current sheet in resistive plasmas[3] and in collisionless plasmas.[4] In space plasmas, collisionless tearing instability might be more relevant, but the process is too slow to be responsible for flares or substorms. In this paper, in connection with explosive phenomena in space plasmas, "magnetic reconnection" is meant to be large-scale physical process in which ambient magnetized plasmas outside X neutral line (point) have significant reconnection flows and magnetic field lines in the ambient magnetized region reconnect at X line. Such large-scale magnetic reconnection was first studied in two-dimensional (2D) steady MHD (magnetohydrodynamic) flows,[5] and the evolutionary process has been studied mainly by MHD and particle simulations. A large number of initial-boundary value problems have been studied for more than forty years, but there is still much confusion regarding physical mechanism of reconnection onset. Hence, main purpose of the present paper is to clarify the fundamental and general feature of large-scale reconnection onset in current sheet in space plasmas without effective Coulomb collision. In general, reconnection rate can be defined as (inductive) reconnection electric field along X neutral line on the basis of Faraday's law. In the growth phase of flare or substorm, magnetic energy may be stored in the form of current sheets by global dynamics of magnetized plasmas in the solar corona or in the geomagnetic tail. So, there are significant inductive electric fields in large-scale region outside X line, where plasmas are magnetized and the frozen-in condition, $\mathbf{E} + \mathbf{u}\times\mathbf{B} = 0$, is satisfied, since plasma bulk flows $\mathbf{u}$ are driven by $\mathbf{J}\times\mathbf{B}$ forces as well as pressure-gradient forces without any influence of stress in electric fields. In this growth phase, large-scale magnetic reconnection must not occur until magnetic energy is critically stored. Hence, large-scale magnetic reconnection may be triggered when inductive electric fields $\mathbf{E}$ ($=-\mathbf{u}\times\mathbf{B}$), which arise as a result of plasma flows $\mathbf{u}$ in the ambient magnetized region, are allowed to penetrate into X line to sustain significant reconnection rate.

## II. PHYSICAL MECHANISM OF LARGE-SCALE RECONNECTION ONSET

An essential question is to clarify physical mechanism of reconnection onset. As a trivial case, in vacuum, magnetic reconnection occurs most easily without any plasma effect. In order to see this, let us consider a 2D case where two rod currents flowing in the z direction are placed at (x=0, y=$\pm$L) in vacuum. If the rod currents flowing in same direction approach each other in the y direction, the origin (x=y=0) becomes an X neutral line. Integrating the y component of Faraday's law, $\nabla \times \mathbf{E} = -\partial \mathbf{B}/\partial t$, from x=0 to $\infty$ along the x axis, we obtain

$$E_z(x=\infty, y=0, t) - E_z(x=0, y=0, t) = d\Phi_y(t)/dt, \qquad (1)$$

where $\Phi_y(t)$ is the integral of $B_y(x, y=0, t)$ from x=0 to $\infty$. Also, integrating the x component of Faraday's law from y=0 to $\infty$ along the y axis, we obtain by taking $E_z(x=\infty, y=0, t) = E_z(x=0, y=\infty, t)=0$

$$E_z(x=0, y=0, t) = d\Phi_x(t)/dt = -d\Phi_y(t)/dt, \qquad (2)$$

where $\Phi_x(t)$ is the integral of $B_x(x=0, y, t)$ from y=0 to $\infty$. Hence, the inductive electric field $|E_z(x=y=0, t)|$ indicates how reconnection proceeds to change overall magnetic field topology. In this respect, note that for any 2D current sheet with antiparallel magnetic fields ($B_x$), usually employed in reconnection studies in plasmas, $|E_z|$ at X neutral line exactly defines the reconnection rate. Hence, an essential question with respect to reconnection onset is to clarify such a physical mechanism that enables $|E_z|$ along X neutral line to arise and be sustained. Apparently, magnetic reconnection in vacuum, which may be called "vacuum reconnection", occurs when ambient inductive electric fields propagate and arrive at X line, and it is completely controlled by external conditions.

The situation is drastically changed when magnetic field lines are embedded in collisioness plasmas. Let us consider the conventional 2D X-type field configuration with antiparallel magnetic fields ($B_x$) and an X neutral line at the origin in the coordinate system same as Eq. (2). In this case, in the ambient large-scale region outside X line, plasmas are magnetized and the frozen-in condition is satisfied, so inductive electric fields can propagate in the form of Alfven waves or magnetosonic waves. On the other hand, there is a small region around X neutral line without any significant influence of magnetic field. In this small region, both ions and electrons are un-magnetized, so when inductive electric fields propagate, they must propagate in the form of electromagnetic waves. As well known, in collisionless plasmas without zero-order magnetic field, the dispersion relation for electromagnetic (transverse) waves in the form of $\exp[i(ky-\omega t)]$ is given by

$$\omega^2 - c^2 k^2 = \omega_{pe}^2, \tag{3}$$

where c and $\omega_{pe}$ are, respectively, light speed and plasma frequency, and only electron dynamics is considered in addition to Maxwell equations. It is well known that as can be expected from Eq. (3), communications with satellite reentering into Earth's atmosphere are often interrupted because of shielding (cut-off) of communication electromagnetic waves around satellite, since the cut-off (plasma) frequency $\omega_{pe}$ around satellite becomes notably enhanced by frictional heating. Eq. (3) indicates that if plasmas are collisionless, ambient inductive electric fields (or electromagnetic waves) with frequency lower than the cut-off frequency $\omega_{pe}$ are shielded (or cut off) around X line in time scale of $1/\omega_{pe}$ by free electrons. Hence, reconnection (inductive) electric fields ($E_z$) with much lower frequency cannot effectively enter nor exist around X line, and length of penetration of reconnection electric field is electron inertial length (or skin depth) $c/\omega_{pe}$ in the limit of $\omega=0$. Note that shielding (or cut-off) of reconnection electric fields around X line is a fundamental and general feature of collisionless (space) plasmas, since Eq. (3) indicates that free un-magnetized electrons are promptly accelerated and cause strong currents around X line so as to shield reconnection electric fields ($E_z$).

In the geomagnetic tail or solar corona, electron inertial length is about 100 times of Debye length and plasmas are almost collisionless. Also, length scale d* of the region where electrons are un-magnetized around X line may be estimated as follows. Let us consider a current sheet around the origin in static equilibrium, where total (magnetic + plasma) pressure is constant, since X neutral line is formed at the origin as a result of reconnection in the current sheet. The associated antiparallel magnetic fields around the origin may be given by $B_{0x}(y)=B_0(y/d)$ in $|y|<d$, $=B_0$ for $y>d$ and $=-B_0$ for $y<-d$, where d is the half thickness of the current sheet and $B_0$ is the ambient antiparallel field strength. Then, the electron gyro-radius based on the magnetic field $B^*=B_0(d^*/d)$ at $y=d^*$ is given by $r_{ge}^* = r_{ge}(d/d^*)$, where $r_{ge} = v_{te}/(eB_0/m_e)$ ($m_e$ is the electron mass and $v_{te}$ the electron thermal velocity). By taking $d^* = r_{ge}^*$, we obtain

$$d^{*2} = r_{ge}\, d . \tag{4}$$

In $|y| < d^*$, electron gyro-radius is larger than $d^*$, so electrons can freely be accelerated in the form of meandering motions by reconnection electric fields $E_z$ ($E_z$ has same sign as sheet current $J_z$). Hence, in the region $|y| < d^*$, electrons can be considered to be un-magnetized, so Eq. (3) is established as follows. In the current sheet, let us consider (reconnection) perturbations in the form of $\exp[i(ky - \omega t)]$, $\mathbf{E} = \mathbf{E}_1$ and $\mathbf{B} = \mathbf{B}_0(y) + \mathbf{B}_1$, where $\mathbf{E}_1 = (0, 0, E_{1z})$ and $\mathbf{B}_0(y) = [B_{0x}(y), 0, 0]$. Then, from Maxwell equations we obtain $\nabla \times \mathbf{E}_1 = -\partial \mathbf{B}_1/\partial t$ and $\nabla \times \mathbf{B}_1 = \mu_0 \mathbf{J}_1 + (1/c^2)\partial \mathbf{E}_1/\partial t$. Here, current density $\mathbf{J} = \mathbf{J}_0 + \mathbf{J}_1$ and $\nabla \times \mathbf{B}_0(y) = \mu_0 \mathbf{J}_0$ [initial sheet current $\mathbf{J}_0 = (0, 0, J_{0z})$, where $J_{0z}$ is constant in $|y| < d$]. Hence, we obtain

$$(\omega^2 - c^2 k^2)E_{1z} = (-i\omega/\varepsilon_0)J_{1z} . \tag{5}$$

In the central un-magnetized region $|y| < d^*$, currents may be caused by electron flow velocity, $\mathbf{u}_e = \mathbf{u}_{e0} + \mathbf{u}_{e1}$, where $\mathbf{u}_{e0} = (0, 0, u_{e0z})$, so we take the current density $J_{0z} = -n_0 e u_{e0z}$ and $\mathbf{J}_1 = -n_0 e \mathbf{u}_{e1}$ ($n_0$ is electron number density). Then, if $B^* = B_{0x}(y=d^*)$ is much smaller than $B_0$, the Lorentz force term, $\mathbf{u}_{e1} \times \mathbf{B}_0(y)$, can be neglected in $|y| < d^*$, where $n_0$ can be taken to be constant since plasma pressure $P_0$ can be considered to be uniform because of $B^{*2} << B_0^2$. Hence, the z component of the linearized equations of electron flow in $|y| < d^*$ becomes

$$m_e\, \partial u_{e1z}/\partial t = -eE_{1z} . \tag{6}$$

Since $J_{1z} = -n_0 e u_{e1z}$, Eqs. (5) and (6) readily give rise to Eq. (3), which indicates that (low-frequency) reconnection electric fields ($E_{1z}$) arising in the ambient magnetized region are shielded in the central un-magnetized region $|y| < d^*$.

In the geomagnetic tail, $r_{ge}$ is comparable with electron inertial length and d is much larger than ion inertial length, so $d^*$ is much smaller than d and much larger than electron inertial length. Hence, inductive (low-frequency) reconnection electric fields ($E_{1z}$) propagating towards the central un-magnetized region (in $|y|<d^*$) in the y direction can hardly penetrate into X line, so (large-scale) magnetic reconnection can hardly occur in usual circumstances. This feature is fundamental for the growth phase of flare or substorm. That is, if there is no reconnection ($E_z=0$ along X line) because of the shielding effect [Eq. (3)], magnetic field lines can be persistently piled up by plasma bulk inflow $u_y$ in the ambient magnetized region, where the frozen-in condition is satisfied. Hence, magnetic energy can be effectively stored in large-scale region in accordance with growth of sheet currents resulting from shielding of ambient reconnection electric fields [Eq. (6)]. In fact, in the geomagnetic tail, it is well known from observations that magnetic energy is critically stored in the form of current sheets before substorm onset. In this respect, note that the shielding effect [Eq. (3)] is established also in the presence of sheared magnetic field ($B_{z0}$) around X line, since z-directional acceleration of free electrons around X line by electric field ($E_{1z}$) is not influenced by sheared fields ($B_{z0}$) in Eq. (6) [the associated (high-frequency) electromagnetic waves are known as "ordinary waves"]. So, in space plasmas without effective Coulomb collision, magnetic energy can be critically stored in the form of current sheets, in general, with sheared fields ($B_{z0}$) because of the shielding effect until explosive reconnection is triggered. Hence, in view of explosive phenomena in space plasmas, "reconnection onset" may be considered to occur when the shielding effect [Eq. (3)] is violated and inductive electric fields in the ambient magnetized region can penetrate into X line to sustain significant reconnection rate. Note that this large-scale reconnection onset is quite different from onset of tearing instabilities that occur locally in current sheet.

The shielding effect [Eq. (3)] indicates that large-scale reconnection cannot occur in (collisionless) space plasmas if there is no anomalous resistivity. So, question is why "collisionless reconnection" occurs in usual particle-in-cell (PIC) simulations,[6] since PIC simulations of collisionless reconnection present large-scale reconnection without anomalous resistivity. So, on the basis of the shielding effect, this question may be interpreted as follows. Large-scale reconnection occurrence implies that the shielding effect is violated around X line numerically or physically, so this must be because PIC simulations take numerical and/or physical parameters in such a way that shielding of reconnection electric fields around X line becomes incomplete. In fact, regarding numerical parameters adopted by usual PIC simulations, number of grids in electron inertial length is forced to be too small, electron inertial length is not sufficiently larger than Debye length, mesh size is not smaller than Debye length, and so forth. Hence, when the shielding is violated physically or numerically, "collisionless reconnection" may occur in the manner similar to "vacuum reconnection", since, for instance, poor numerical resolution inside electron inertial length (or skin depth) around X line allows ambient reconnection electric fields to penetrate into X line to cause "fast" reconnection without physical dissipation as in vacuum. In fact, so-called global PIC simulations with much lower numerical resolution cannot involve the shielding effect around X line and show that "collisionless reconnection" occurs more easily, despite that physical quantities like reconnection electric field cannot precisely be measured inside electron inertial length around X line because of coarse grids.[7] This may provide an evidence that collisionless reconnection occurs because the shielding effect does not work because of poor numerical resolution. Also, recent PIC simulations with standard numerical parameters demonstrate that "collisionless reconnection" occurs at faster rate for worse numerical resolution.[8] Namely, Fig. 2 in Ref. 8 indicates that the reconnection rate in the main phase, subsequent to the peak value that is directly caused by artificial initial perturbation, becomes remarkably smaller for higher numerical resolution (with larger number of grid points in initial current sheet thickness) with the other conditions except for the domain size being the same; in this simulation, open boundaries are taken, so the domain size has no significant effect on plasma processes occurring in the domain. These results cannot be explained by any plasma effect around X line, since any physical effect inside the un-magnetized region $|y|<d^*$ (or inside electron inertial length) around X line becomes less effective for worse numerical resolution. The only answer is that shielding of reconnection electric fields by collisionless electrons around X line [Eq. (3)] becomes less effective for worse numerical resolution, leading to faster reconnection. This is apparently because the shielding effect around X line has never been noticed in PIC simulations, so no one has recognized that physical quantities in $|y|<d^*$ (even inside electron inertial length) must be precisely analyzed. Also, note that large currents seen around X line in some recent PIC simulations (e.g., Ref. 8) may result from the shielding effect by free electrons around X line [Eq. (6)]. But, the shielding may be still incomplete because of lack of numerical resolution around X line, so some ambient reconnection electric fields are allowed to penetrate into X line to cause collisionless reconnection. Of course, collisionless reconnection may be a physical process occurring in some physical situations, since, for instance, non-zero (very small) reconnection electric fields could exist at X line because of electron inertial effect in thin current sheets even if the shielding effect [Eq. (3)] is complete, although the associated reconnection process should be too slow to be applied to flares or substorms.

It has often been argued that Hall effects or electron anisotropic pressure effects are important for onset mechanism of (large-scale) collisionless reconnection.[6] However, in collisionless plasmas, these effects may be negligibly small inside electron inertial length around X neutral line, where electrons are un-magnetized. Hence, if electron inertial length is much larger than Debye length, shielding of reconnection electric fields around X line [Eq. (3)] can hardly be violated by these effects. Also, it is frequently suggested that effective resistivity due to electron anisotropic pressure gradients or so-called "inertial resistivity" may drive (large-scale) collisionless reconnection.[9] However, note that in order for inductive (reconnection) electric field (due to $\partial \mathbf{B}/\partial t$) to arise at X line in initial current sheet, current density distribution around X line must be suddenly changed. Apparently, drastic change in current density distribution around X line can hardly be caused by such (collisionless) inertial resistivity, so it cannot drive any effective (large-scale) magnetic reconnection in initial current sheet. In this respect, note that the generalized Ohm's law indicates plasma effects that can support electric fields which are assumed to exist steadily. Hence, it is not surprising that these effects, like electron anisotropic pressure effects, exist near X line when reconnection electric fields arise steadily around X line physically or numerically.

Regarding MHD reconnection studies with Ohm's law $\mathbf{E}+\mathbf{u}\times\mathbf{B}=\eta\mathbf{J}$, situations and treatments are much simpler. For collisionless plasmas, resistivity $\eta$ is set to be zero, whereas for collisional plasmas, resistivity models are mathematically formulated according to physical mechanisms causing effective particle collision. For collisionless plasmas with zero resistivity, electric field along X neutral line remains to be always zero, so magnetic reconnection cannot occur. When plasmas become collisional with nonzero resistivity, shielding of reconnection electric field around X line [Eq. (3)] is violated, so reconnection occurs. This situation can be most clearly seen from 2D MHD simulations by Ugai and Tsuda.[10] They assumed localized anomalous resistivity around an X line, so the localized region, where shielding of reconnection electric fields is violated, works like a "hole" where ambient reconnection electric fields can enter. Initially, there is a current sheet with antiparallel magnetic fields $B_x(y)$ (in the coordinate system same as before) in static equilibrium. When anomalous resistivity is locally enhanced in the current sheet, current density is suddenly reduced locally around the X line because of magnetic field diffusion to cause inductive (reconnection) electric field $E_z$ there, but the resulting "local reconnection ($|E_z|$ at the X line)" soon decays because of decrease in current density at X line. In this initial phase, there is no inductive electric field $E_z$ in the ambient magnetized region outside the X line, since there is no plasma bulk flow. They found that large-scale reconnection evolves in the following manner. Reconnection (or $E_z$) at the X line merely causes changes in magnetic fields locally around the X line on the basis of Faraday's law, giving rise to **J**×**B** forces to accelerate reconnection plasma bulk flows **u**. In the ambient magnetized region outside the X line, where the frozen-in condition is satisfied, plasma flows **u** are determined by MHD momentum equations involving **J**×**B** and pressure-gradient forces, and inductive (reconnection) electric field $E_z$ (=$u_y B_x$) grows as a result of growing reconnection inflow $u_y$. When $E_z$ in the magnetized region propagates in the form of magnetosonic waves and is carried inwards by the reconnection inflow $u_y$, it can penetrate into the X line, so reconnection rate ($|E_z|$ at the X line) is enhanced, which gives rise to further magnetic field diffusion around the X line and enhances **J**×**B** forces to further accelerate plasma flows. In this way, the self-consistent interaction eventually leads to fast reconnection mechanism involving slow shocks in the ambient magnetized region. Note that the large-scale magnetic reconnection drastically evolves only after reconnection electric fields, which grow with growing reconnection flows in the ambient magnetized region, effectively penetrate into the X line. Although the fixed resistivity model is artificial, these results indicate underlying physics of fast reconnection evolution. In fact, important effects of localized anomalous resistivity on fast reconnection (Petschek) mechanism have been theoretically recognized.[11]

On the basis of results by Ugai and Tsuda, physical mechanism of fast reconnection evolution has been studied by 2D MHD simulations in the following manner.[12] As well known, kinetic current-driven instabilities give rise to effective anomalous resistivities;[13] in particular, ion-acoustic and Buneman instabilities may well occur around X line.[14] On the basis of well-known features of anomalous resistivities, in the anomalous resistivity model, resistivity $\eta(\mathbf{r}, t)$ is given by function of relative electron-ion drift velocity $|\mathbf{V_D}(\mathbf{r},t)|=|\mathbf{J}(\mathbf{r}, t)|/\rho(\mathbf{r},t)$ ($\rho$ is plasma density) or current density $|\mathbf{J}(\mathbf{r},t)|$ as follows: for $|\mathbf{V_D}(\mathbf{r},t)|$ (or $|\mathbf{J}(\mathbf{r},t)|$) > $V_C$ (or $J_C$), $\eta(\mathbf{r}, t)$ increases with $|\mathbf{V_D}(\mathbf{r},t)|$ (or $|\mathbf{J}(\mathbf{r},t)|$), and for $|\mathbf{V_D}(\mathbf{r},t)|$ (or $|\mathbf{J}(\mathbf{r},t)|$) < $V_C$ (or $J_C$), $\eta=0$ [$V_C$ (or $J_C$) is the threshold given by function of temperature (thermal velocity)]. Hence, $\eta$ is determined self-consistently with temporal plasma dynamics and arises only when and where the threshold is exceeded. Also, initial disturbance is given by a localized resistivity only in short time range in initial current sheet in static equilibrium; after the disturbance is removed, $\eta$ is zero everywhere until the threshold is exceeded. Simulations have been done for a wide range of parameters of the resistivity model and the threshold for different functional forms of anomalous resistivity with free (open) boundaries enclosing simulation domain. It was demonstrated in any case that initiated by disturbance, plasma bulk flows grow so as to cause distinct increase in current density locally around X line, so current-driven anomalous resistivities are eventually driven and sustained locally around X line; once anomalous resistivities arise around X line, the shielding effect is violated there, so reconnection flows and anomalous resistivities drastically grow spontaneously as a sort of nonlinear instability, eventually leading to (Petschek-like) fast reconnection mechanism involving standing slow shocks. Hence, the resulting reconnection evolution is called "spontaneous fast reconnection". It was also shown with the same initial disturbance that reconnection evolution is strongly controlled by resistivity model; for Spitzer resistivity due to Coulomb collision, given by function of plasma temperature, or for uniform resistivity, any slow shock cannot be caused, so no fast reconnection mechanism can be realized.[12]

In order to exemplify the most basic structure of fast reconnection evolution, Fig. 1 shows the magnetic field configuration with current density distribution that has been obtained as an eventual solution from a 2D MHD simulation of the spontaneous fast reconnection model.[12] Here, the conventional symmetry boundary is imposed on the y axis, and inflow and outflow boundaries enclosing the simulation domain are free (open) boundaries. Initially, there was a current sheet with antiparallel magnetic fields ( $B_x$ ) with zero resistivity in static equilibrium. Initiated by a disturbance around the origin, plasma flows grew so as to cause drastic enhancement of current density locally around the origin (X line). In due course, current-driven anomalous resistivities were driven and sustained locally around the X line self-consistently with growth of global reconnection flows, which enabled ambient reconnection electric fields ( $E_z$ ) to effectively penetrate into the X line since the shielding effect was violated around the X line. Hence, the fast reconnection mechanism shown in Fig. 1 has been realized as an eventual solution of MHD equations. Note that plasma bulk flow velocities have spontaneously grown because of **J**×**B** forces resulting from proceeding of magnetic reconnection without any specific external condition. Attached to the small diffusion region around the X line, a pair of (Petschek-like) slow (switch-off) shocks, indicated by strong current layers in Fig. 1, are caused in $|x|<X_1$ and extending in the positive and negative x directions, so the reconnection outflow jet between the pair of slow shocks exactly attains the Alfven velocity $V_A$ measured in the upstream (low-$\beta$) region. Ahead of the Alfvenic reconnection jet, a large-scale plasmoid is formed in the extent $X_1<x<X_2$ and swells according to proceeding of reconnection, so $X_1$ and $X_2$ increase with time. Hence, the fast reconnection process, initially triggered locally around the origin, is rapidly expanding outwards through MHD waves, so the field configuration in $x>X_2$ is not yet notably affected and remains to be almost the initial antiparallel one. In this fast reconnection configuration, magnetic energy initially stored in ambient antiparallel fields ( $B_x$ ) can explosively be released by standing slow shocks as was first recognized by Petschek.[5] Note that this basic fast reconnection configuration (Fig. 1) is not affected by choice of parameters of current-driven anomalous resistivity model.[12] The spontaneous fast reconnection model has extensively been studied by three-dimensional (3D) MHD simulations, where magnetic neutral line has finite length.[15] In 3D situations, only when neutral line length is sufficiently larger than current sheet thickness, fast reconnection mechanism can be realized. Details of the 3D reconnection evolution was recently summarized.[16]

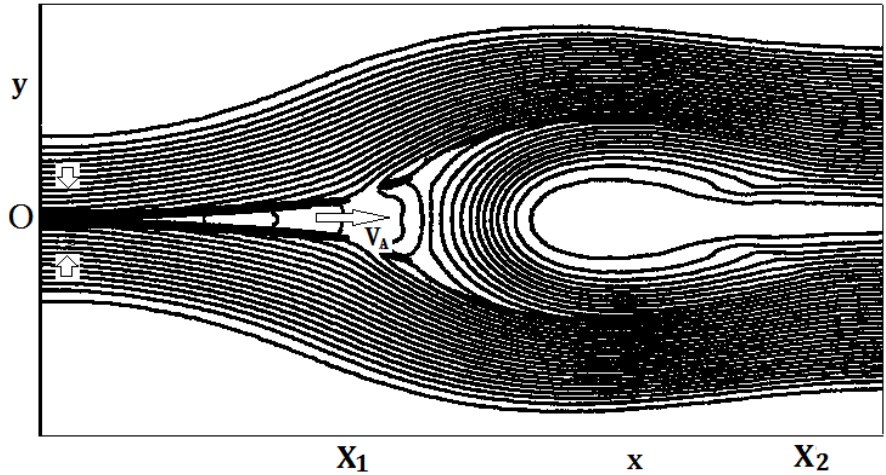

FIG. 1. Magnetic field configuration with current density distribution obtained as an eventual solution from a 2D MHD simulation of the spontaneous fast reconnection model, where an X neutral line is located at the origin O, and white arrows indicate resulting reconnection inflow and outflow velocities. Because of the conventional symmetry conditions, only the right half is shown.

    The spontaneous fast reconnection model has been successfully applied to

well-known characteristic phenomena observed in flares or substorms. For instance, large-scale magnetic loop like flaring loop is formed when the Alfvenic fast reconnection outflow jet (Fig. 1) collides with an obstacle, and a fast shock is formed transiently ahead of the loop top.[17] In 3D situations, so-called current wedge as well as generator current circuit with field-aligned currents along loop boundary is formed, which is generally consistent with two-ribbon flaring heating in loop foot-points or breakup of substorms.[18] Also, the fast reconnection evolution in long current sheet gives rise to large-scale plasmoids (Fig. 1) propagating in both directions from X neutral line.[19] Note that plasmoid propagation can be directly detected by *in situ* satellite observations in the geomagnetic tail.[20] It was demonstrated that 3D plasmoid structure resulting from the spontaneous fast reconnection model in weakly sheared current sheets is, both qualitatively and quantitatively, in good agreement with satellite observations.[21] In addition, the fast reconnection evolution in an initially force-free current sheet can cause extreme coronal heating since temperature becomes almost $1/\beta$ times larger in the fast reconnection jet and plasmoid regions,[22] low-frequency MHD waves like Pi-2 pulsations associated with substorms result from collision between fast reconnection jet and magnetic loop,[23] and so forth.

## III. CONCLUDING REMARKS

We have argued for the first time that large-scale reconnection occurs in space plasmas only when ambient inductive (low-frequency) reconnection electric fields are allowed to effectively penetrate into X neutral line. In usual circumstances, shielding (cut-off) of reconnection electric fields around X neutral line [Eq. (3)] prevents reconnection from occurring effectively and causes critical storage of magnetic energy in large-scale region in the form of current sheets. Hence, when the shielding effect around X line is violated and becomes incomplete, magnetic reconnection can occur, so it is essential for any (MHD or particle) reconnection study to clarify the physical mechanism by which the shielding effect around X line is violated. In fact, in the solar corona or geomagnetic tail without effective Coulomb collision, electron inertial length is much larger than Debye length, so shielding of reconnection electric fields around X line may be almost complete, and magnetic reconnection can hardly occur in usual circumstances. Hence, reconnection should be triggered when plasmas become effectively collisional and dissipative around X neutral line, since then shielding of reconnection electric fields is violated. We have hence proposed that current-driven anomalous resistivities due to wave-particle collisions around X line are essential for fast reconnection evolution responsible for substorms and flares. In this respect, it is often argued that size of diffusion region with such anomalous resistivities must be unrealistically small compared to scale of solar corona. But, note that the localized diffusion region simply works like such a hole that allows ambient reconnection electric fields to penetrate into X line, so however small, it can work effectively. In fact, in laboratory reconnection experiments with high-temperature plasmas, current-driven anomalous resistivities are detected in an extremely small region around X neutral line.[24] Also, in the geomagnetic tail, it is demonstrated with Cluster observations that current-driven anomalous resistivities play dominant role in magnetic energy conversion.[25]

In summary, the main theme of the present paper is to propose a new physical concept on reconnection onset in current sheet and to clarify physical mechanism of explosive magnetic energy conversion responsible for flares and substorms observed in space plasmas. As well known, the most plausible physical mechanism of explosive magnetic energy conversion is the (Petchek-like) fast reconnection mechanism involving standing slow shocks, since slow shock is the most powerful magnetic energy converter in space plasmas. So far, it is only the spontaneous fast reconnection model that can demonstrate drastic evolution of the (Petschek-like) fast reconnection mechanism involving slow shocks as an eventual solution (Fig. 1). In order to understand the underlying physical mechanism of the fast reconnection evolution, we propose that large-scale reconnection can occur when the shielding effect around X line [Eq. (3)] is violated and ambient reconnection (inductive) electric fields effectively penetrate into X neutral line. This concept can reasonably explain the physical mechanism of reconnection onset from vacuum to MHD in a unified manner. Obviously, the physical concept on large-scale reconnection onset in the spontaneous fast reconnection model is quite different from that in the conventional reconnection model. In the spontaneous fast reconnection model, local plasma conditions around X line are merely changed so as to allow reconnection (inductive) electric fields arising in the ambient magnetized region (due to **u**×**B** term) to penetrate into X line, which enables magnetic field lines in the ambient magnetized region to reconnect at X line; hence, it is large-scale dynamics of magnetized plasmas outside X line that plays the dominant role in explosive conversion of magnetic energy into plasma energies.

# REFERENCES

[1] E. R. Priest, Rep. Prog. Phys. **48**, 955, DOI: 10.1088/0034-4885/48/7/002 (1985); A. T. Y. Lui, Space Sci. Rev. **95**, 325 (2001).

[2] E. R. Priest and T. G. Forbes, *Magnetic reconnection: MHD Theory and Applications* (Cambridge University Press, UK, 2000); M. Scholer, J. Geophys. Res. 94, 8805, DOI:10.1029/JA094iA07p08805 (1989); J. A. Klimchuk, Sol. Phys. **234**, 41 (2006); K. Shibata, Astrophys. Space Sci. **264**, 129 (1999).

[3] H. P. Furth, J. Killeen, and M. N. Rosenbluth, Phys. Fluids **6**, 459 (1963).

[4] B. Coppi, G. Laval, and R. Pellat, Phys. Rev. Lett. **16**, 1207 (1966).

[5] H. E. Petschek, *NASA Special Publication 50* (National Aeronautics and Space Administration, Washington, DC, 1964), p. 425.

[6] J. Birn, J. F. Drake, M. A. Shay, B. N. Rogers, R. E. Denton, M. Hesse, M. Kuznetsova, Z. W. Ma, A. Bhattacharjee, A. Otto, and P. L. Pritchett, J. Geophys. Res. **106**, 3715, DOI:10.1029/1999JA900449 (2001).

[7] D. S. Cai, W. Tao, X. Yan, B. Lembege, and K-I. Nishikawa, J. Geophys. Res. **114**, A12210, DOI: 10.1029/2007JA012863 (2009).

[8] H. E. Sun, Z. W. Ma, and J. Huang, Phys. Plasmas **21**, 072115, DOI: 10.1063/1.4889894 (2014).

[9] N. Singh, Phys. Plasmas **21**, 030704, DOI: 10.1063/1.4869723 (2014); M. Yamada, R. M. Kulsrud, and H. Ji, Rev. Modern Phys. **82**, 603, DOI: 10.1103/RevModPhys.82.603 (2010).

[10] M. Ugai and T. Tsuda, J. Plasma Phys. **17**, 337 (1977); **22**, 1 (1979).

[11] R. M. Kulsrud, Earth Planets Space **53**, 417 (2001); L. M. Malyshkin, T. Linde, and R. M. Kulsrud, Phys. Plasmas **12**, 102902, DOI: 10.1063/1.2084847 (2005); T. G. Forbes, E. R. Priest, D. B. Seaton, and Y. E. Litvinenko, Phys. Plasmas **20**, 052902, DOI:10.1063/1.4804337 (2013).

[12] M. Ugai, Plasma Phys. Controlled Fusion **26**, 1549 (1984); **27**, 1183 (1985); Phys. Fluids **29**, 3659 (1986); Phys. Fluids B **4**, 2953 (1992); Phys. Plasmas **6**, 1522 (1999).

[13] A. T. Y. Lui, Space Sci. Rev. **113**, 127 (2004); R. A. Treumann, Earth, Planets Space **53**, 453 (2001).

[14] P. Petkaki and M. P. Freeman, Astrophys. J., 686, 686 (2008).

[15] M. Ugai, Phys. Plasmas **15**, 082306 (2008).

[16] M. Ugai, Phys. Plasmas **19**, 072315 (2012).

[17]M. Ugai, Geophys. Res. Lett. **14**, 103, DOI: 10.1029/GL014i002p00103 (1987); Phys. Plasmas **3**, 4172 (1996).

[18]M. Ugai, Phys. Plasmas **15**, 032902 (2008); **16**, 012901 (2009).

[19]M. Ugai, Phys. Plasmas **2**, 3320 (1995).

[20]J. A. Slavin, R. P. Lepping, J. Gjerloev, D. H. Fairfield, M. Hesse, C. J. Owen, M. B. Moldwin, T. Nagai, A. Ieda, and T. Mukai, J. Geophys. Res. **108**, 1015 (2003); J. A. Slavin, M. F. Smith, E. L. Mazur, D. N. Baker, E. W. Hones Jr., T. Iyemori, E. W. Greenstadt, J. Geophys. Res. **98**, 15425 (1993).

[21]M. Ugai, J. Atmos. Solar-Terrestrial Phys. **99**, 47 (2013); Ann. Geophys. **29**, 1411 (2011).

[22]M. Ugai, Phys. Plasmas **18**, 102903 (2011).

[23]M. Ugai, Phys. Plasmas **16**, 112902 (2009).

[24]Y. Ono, M. Inomoto, Y. Ueda, T. Matsuyama, and Y. Murata, Earth Planets Space, **53**, 521 (2001); H. Ji, M. Yamada, S. Hsu, and R. Kulsrud, Phys. Rev. Lett. **80**, 3256 (1998).

[25]A. T. Y. Lui, Y. Zheng, H. RèMe, M. W. Dunlop, G. Gustafsson, and C. J. Owen, J. Geophys. Res. **112**, A04215, DOI: 10.1029/2006JA012000 (2007).